\documentclass[aps,pra,reprint,superscriptaddress,footinbib]{revtex4-2}

\makeatletter
\renewenvironment{widetext}{%
  \par\ignorespaces
  \setbox\widetext@top\vbox{}%
  \setbox\widetext@bot\hbox{}%
  \onecolumngrid
  \vskip 10\p@
  \dimen@\ht\widetext@top
  \advance\dimen@\dp\widetext@top
  \cleaders\box\widetext@top\vskip\dimen@
  \vskip 6\p@
  \prep@math@patch
}{%
  \par
  \vskip 6\p@
  \setbox\widetext@bot\vbox{}%
  \dimen@\ht\widetext@bot
  \advance\dimen@\dp\widetext@bot
  \cleaders\box\widetext@bot\vskip\dimen@
  \vskip 8.5\p@
  \twocolumngrid
  \global\@ignoretrue
  \@endpetrue
}
\makeatother

\usepackage[T1]{fontenc}				
\usepackage[utf8]{inputenc}			
\usepackage[english]{babel}			
\usepackage[protrusion=false]{microtype}	

\usepackage{xcolor} 						
\definecolor{red}{rgb}{1,0,0}				
\definecolor{blue}{rgb}{0,0,1}				
\definecolor{black}{rgb}{0,0,0}				

\usepackage{soul} 
\definecolor{hlyellow}{rgb}{0.95,0.95,0}
\definecolor{hlgreen}{rgb}{0,0.95,0}
\sethlcolor{hlyellow}
\soulregister \ref 7
\soulregister \cite 7
\soulregister \citet 7
\soulregister \footnote 8
\soulregister {\ } 7
\soulregister \figref 7
\soulregister \tabref 7
\soulregister \secref 7
\soulregister \appref 7
\soulregister \refref 7
\soulregister \eqref 7
\soulregister \supmat 7

\usepackage{amsmath} 
\usepackage{amssymb,amsfonts,mathrsfs} 
\usepackage{array} 
\usepackage{upgreek} 
\usepackage{braket}
\usepackage{bbold}

\usepackage{graphicx} 
\usepackage{makecell}
\usepackage{longtable}
\usepackage{booktabs}
\usepackage{placeins} 
\usepackage{svg}

\usepackage{comment}
\usepackage{hyperref} 
\definecolor{dullmagenta}{rgb}{0.4,0,0.4} 
\definecolor{darkblue}{rgb}{0,0,0.4}
\definecolor{medblue}{rgb}{0,0,0.6}
\definecolor{lightblue}{rgb}{0,0,0.8}
\hypersetup{
	colorlinks=true, 		
	breaklinks=true,		
	linkcolor=lightblue,
	citecolor=lightblue,
	urlcolor=lightblue,
	filecolor=dullmagenta
}
\usepackage[all]{hypcap} 

\newcommand{\figletter}[1]{\textbf{(\mbox{#1})}}
\newcommand{\figref}[1]{Fig.~\ref{#1}} 
\newcommand{\figreflet}[2]{Fig.~\ref{#1}(#2)} 
\newcommand{\figrefs}[2]{Figs.~\ref{#1}-\ref{#2}} 
\newcommand{\tabref}[1]{Tab.~\ref{#1}} 
\newcommand{\secref}[1]{Sec.~\ref{#1}} 
\newcommand{\appref}[1]{Appendix \ref{#1}} 
\newcommand{\refref}[1]{Ref.~\cite{#1}} 

\newcommand{\GeSiGe}{Ge/SiGe} 

\newcommand{\V}{\ \text{V}}					
\newcommand{\mV}{\ \text{mV}}				
\newcommand{\uV}{\ \upmu\text{V}}				
\newcommand{\uA}{\ \upmu\text{A}}				
\newcommand{\ueV}{\ \upmu\text{eV}}				
\newcommand{\GHz}{\ \text{GHz}}				
\newcommand{\um}{\ \upmu\text{m}}		
\newcommand{\nm}{\ \text{nm}}			
\newcommand{\nA}{\ \text{nA}}			
\newcommand{\dBm}{\ \text{dBm}}			

\newcommand{\mK}{\ \text{mK}}	

\newcommand{\supmat}{SI} 

\newcommand{\SupRefFigStat}[1]{\figref{supfig:stat}}

\newcommand{\SupSecMar}[1]{\ref{sec:sup_mar}}
\newcommand{\SupSecQPC}[1]{\ref{sec:additional_QPCs}}
\newcommand{\SupSecStat}[1]{\ref{sec:additional_QPCs}}

\begin{document}

\title{A superconducting quantum point contact in Ge}

\newcommand{\zrl}{IBM Research Europe -- Zurich, Säumerstrasse 4, 8803 Rüschlikon, Switzerland}
\newcommand{\yorktown}{IBM Research, T.\,J.\,Watson Research Center, 1101 Kitchawan Road, Yorktown Heights, New York 10598, USA}

\newcommand{\equal}{These authors equally contributed to this work.}

\author{Anna~\surname{Tacchini}}
\thanks{These authors contributed equally to this work.}
\affiliation{\zrl}

\author{Inga~\surname{Seidler}}
\thanks{These authors contributed equally to this work.}
\affiliation{\zrl}

\author{Isabell~\surname{Grandt-Ionita}}
\affiliation{\zrl}

\author{Thibault~\surname{Thuegaz}}
\affiliation{\zrl}

\author{Stephen~W.~\surname{Bedell}}
\affiliation{\yorktown}

\author{Fabrizio~\surname{Nichele}}
\affiliation{\zrl}

\begin{abstract}
Hybrid superconducting-semiconducting circuits provide a versatile platform for investigating coherent transport in mesoscopic weak links. Superconducting quantum point contacts (SQPCs) are particularly interesting in this context, as their supercurrent is carried by a discrete number of ballistic transport modes. Here, we investigate a gate-tunable superconducting quantum point contact realized in Ge/SiGe. By incorporating the QPC into a highly asymmetric superconducting quantum interference device (SQUID), we gain access to its current-phase relation (CPR) and switching current, enabling phase-sensitive studies of few-mode superconducting transport. We observe a direct correspondence between conductance quantization and switching-current quantization, linking the normal-state transport modes to the superconducting response. In addition, we demonstrate strong gate tunability of the effective transmission near the onset of the first conductance plateau. Under microwave irradiation, the QPC exhibits photon-assisted multiple Andreev reflections at lower transparencies and Shapiro replicas in the high-transparency, few-mode regime.

\end{abstract}

\date{12 November 2026}

\maketitle

\section{Introduction}\label{sec1}
Andreev bound states (ABSs), discrete subgap excitations that carry supercurrent through short Josephson weak links~\cite{Andreev_thermal_1964,beenakker_josephson_1991,furusaki_current-carrying_1991}, have attracted growing interest for their role in topological superconductivity~\cite{Lutchyn2018Majorana, prada_andreev_2020},  and for implementing qubits~\cite{desposito_controlled_2001,Zazunov_Andreev_2003,lee_spin-resolved_2014,janvier_coherent_2015,aguado_perspective_2020,hays_coherent_2021,pita-vidal_direct_2023}. 
Controlling these states also enables new superconducting circuit functionalities, including Josephson diodes~\cite{souto_josephson_2022,ciaccia_gate_2023,matsuo_josephson_2023,valentini_parity-conserving_2024,coraiola_flux_2024} and circuit elements with tailored current–phase relations (CPRs)~\cite{spanton_current_2017,leblanc2025gate, banszerus_hybrid_2025}. These prospects motivate the development of material platforms that enable control over the number and transmission probabilities of the conducting channels supporting ABSs. Ballistic superconducting transport has been thoroughly investigated in atomic break junctions~\cite{goffman_supercurrent_2000,ludoph_multiple_2000,della_rocca_measurement_2007, bretheau_exciting_2013} and gate-defined superconducting quantum point contacts (SQPCs)~\cite{takayanagi_observation_1995,takayanagi_andreev_1995,irie_josephson_2014,mikheev_quantized_2021}.
Controlled occupation of discrete electronic modes produces steps in the critical current~\cite{beenakker_josephson_1991, beenakker_superconducting_1992}, analogous to conductance quantization in normal-state quantum point contacts (QPCs).

\GeSiGe{} heterostructures have emerged as a promising platform for hybrid superconductor–semiconductor devices~\cite{vigneau_germanium_2019,hendrickx_ballistic_2019,aggarwal_enhancement_2021,tosato_hard_2023,valentini_parity-conserving_2024,sagi_gate_2024,lakic_quantum_2025,hinderling_direct_2024,ten_kate_finite-length_2025,borovkov_low-noise_2026,ruggiero_multi-level_2026,viswanathan_buried_2026,sokolova_superconducting_2026},
combining low disorder, strong spin–orbit interaction and highly transparent contacts~\cite{scappucci_germanium_2021}.  These properties support gate-defined devices hosting a few highly transmitting channels, as demonstrated by recent observations of Andreev-enhanced conductance quantization~\cite{kiyooka_andreev-enhanced_2026,sokolova_superconducting_2026}. Recent work also investigated the compatibility of \GeSiGe{} with microwave techniques~\cite{valentini_parity-conserving_2024,hinderling_direct_2024,ten_kate_finite-length_2025, palma_low-loss_2025,ruggiero_multi-level_2026}, which provides a formidable tool for probing and controlling Andreev-bound states~\cite{chauvin_superconducting_2006,fuechsle_effect_2009,virtanen_theory_2010,bergeret_supercurrent_2011, haxell_microwave-induced_2023,costa_sign_2023}.

Precise control over individual channels could enable engineering of Andreev spectra in more complex Josephson architectures, including multiterminal devices~\cite{riwar2016multiterminal,pankratova_multiterminal_2020,coraiola_phase-engineering_2023}, non-Hermitian devices~\cite{beenakker_nonhermitian_2024,cayao_non-hermitian_2024,li_anomalous_2024,shen_non-hermitian_2024} and novel qubit architectures based on a single highly transmissive Andreev state~\cite{matute-canadas_quantum_2024,tjernshaugen2026coherentcontrolspinmons,manesco_loopless_2026,del_olmo_n_inductively-protected_2026,caceres_ferbo_2026}.
Moreover, \GeSiGe{} is an established spin-qubit platform~\cite{scappucci_germanium_2021}, offering opportunities to integrate superconducting channels with quantum-dot circuits~\cite{leijnse_coupling_2013,jakob2025fastreadoutquantumdot,lakic_quantum_2025, borovkov_low-noise_2026}.

Here, we investigate an electrostatically defined \GeSiGe{} SQPC embedded in a highly asymmetric superconducting quantum interference device (SQUID). This geometry gives access to the CPR, allowing us to characterize how the modes identified through the normal state conductance quantization carry supercurrent at different superconducting phases. We observe corresponding conductance and switching-current plateaus, together with wide gate tunability of both the supercurrent amplitude and effective channel transmissions. We discuss how non-unity transmission, thermal ABS occupation, and switching dynamics can contribute to deviations from the ideal short-junction response. Microwave irradiation probes ABS dynamics and offers a complementary means of controlling the superconducting response. We observe photon-assisted multiple Andreev reflections and Shapiro replicas in different gate regimes, as well as microwave-induced CPR changes beyond the adiabatic response, reflecting the occupation of excited Andreev states. In light of these results, \GeSiGe{} SQPCs offer a tunable building block for engineering superconducting circuits at the level of a few quantum channels.

\begin{figure*}[htbp]
    \centering
    \includegraphics[width=\textwidth]{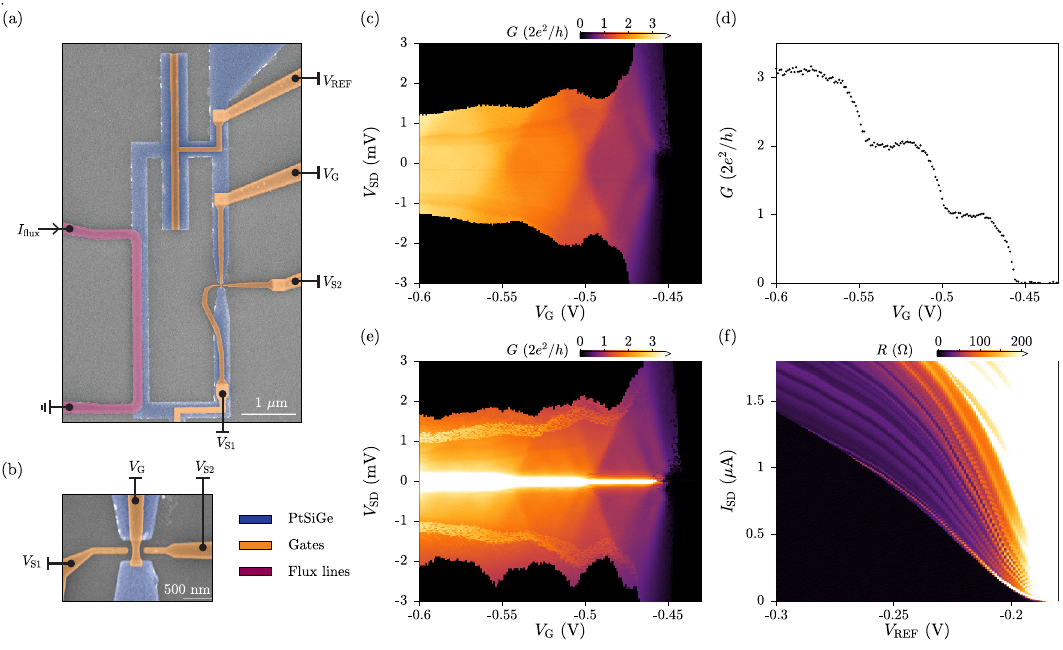}
    \caption{Device geometry and individual JJ characteristics of device A.  
    \figletter{a}~Colored scanning electron micrograph of a device identical to the one under study with a reference junction (controlled by $V_\mathrm{REF}$) and SQPC (controlled by $V_\mathrm{G}$) and superconducting flux line (purple). 
    \figletter{b}~Zoom of panel (a) focusing on the SQPC. \figletter{c}~Differential conductance as a function of gate voltage $V_\mathrm{G}$ and source-drain bias $V_\mathrm{SD}$ of the SQPC  at $T=700\mK$ and $V_\mathrm{REF}=0\V$. \figletter{d}~Differential conductance of the SQPC  at $T=700\mK$, $V_\mathrm{REF}=0\V$ and $V_\mathrm{SD}=0\V$.  \figletter{e}~Differential conductance as a function of the gate voltage $V_\mathrm{G}$ and source-drain bias $V_\mathrm{SD}$ of the SQPC  at $T=7\mK$ and $V_\mathrm{REF}=0\V$. \figletter{f}~Differential resistance of the reference junction as a function of bias current $I_\mathrm{SD}$ and gate voltage $V_\mathrm{REF}$ at $V_\mathrm{G}=0\V$.}
    \label{fig:device}
\end{figure*}

\section{Results and discussion}\label{sec2}

\subsection{Individual Junction Characterization}
The SQUID consists of two superconductor–normal semiconductor–superconductor (SNS) JJs.  The JJs are realized by contacting the quantum well with superconducting PtSiGe \cite{tosato_hard_2023}. A false-colored scanning electron image of an identical device is depicted in \figreflet{fig:device}{a}. The superconducting loop, highlighted in blue, is formed from PtSiGe, while the electrostatic gates, shown in yellow, control the two junctions. A flux line enables control of the SQUID phase. The large JJ, with length $500\nm$ and width $1.4\um$, controlled by gate voltage $V_\mathrm{REF}$, is referred to as the reference junction. The QPC has a length of $500\nm$ and a gate width of $100\nm$, and is controlled by gate voltage $V_\mathrm{G}$.  Its magnified view is shown in~\figreflet{fig:device}{b}.  

First, the SQPC and the reference junction are examined separately. Source–drain bias spectroscopy was performed on the SQPC at $700\mK$, well above the critical temperature of the superconducting PtSiGe, by applying a DC bias voltage ($V_{\mathrm{SD}}$) and measuring the differential conductance G (see Methods). Three conductance plateaus were observed (\figreflet{fig:device}{c}). The zero-bias conductance plateaus correspond to integers of $2e^2/h$ (\figreflet{fig:device}{d}). At finite bias odd integer multiples of $e^2/h$ plateaus appear. The extracted subband energy spacing is $\Delta E=2~\mathrm{meV}$. 
At base temperature, a pronounced zero-bias conductance enhancement is observed (\figreflet{fig:device}{e}), providing evidence of superconducting transport. The measured conductance greatly exceeds its normal-state value, reaching values above $50\cdot 2e^2/h$, yet remains finite. This prevents a reliable determination of the junction's switching current from its current–voltage characteristics alone, most likely due to a high  phase escape rate at zero-bias. This limitation motivates the use of an alternative measurement technique to reliably determine the junction's switching current. At biases above the induced superconducting gap, quasiparticle transport dominates and the conductance approaches its normal-state value.

\figreflet{fig:device}{f} shows the differential resistance of the reference junction as a function of bias current and gate voltage. The switching current of the reference junction could be continuously tuned by $V_\mathrm{REF}$ from zero to $1.5\uA$, whereas the normal state resistance was approximately $20\,\Omega$.

\begin{figure}[tbp]
    \centering
    \includegraphics[width=\columnwidth]{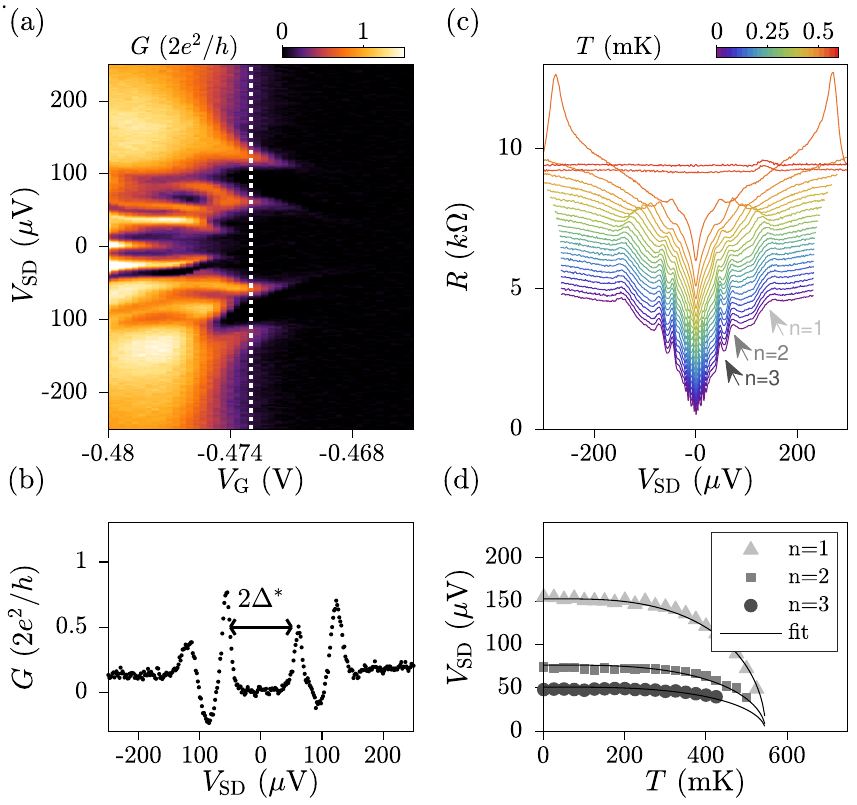}
    \caption{Multiple Andreev reflection of the SQPC of the device A.  
    \figletter{a}~Differential conductance of the SQPC at the onset of the first conductance plateau of the SQPC. 
    \figletter{b}~Line cut of panel a at $V_\mathrm{G}=-0.472\V$ (black dots) with marked conductance peaks separation $2\Delta^*$. \figletter{c}~Differential resistance of the SQPC at $V_\mathrm{G}=-0.6\V$ as a function of source-drain voltage $V_\mathrm{SD}$ and temperature $T$. Each temperature curve is offset by $200~\Omega$. The peaks shifting with temperature indicated by arrows correspond to low-order MAR processes. \figletter{d}~Evolution of the three MAR peaks as a function of the temperature and voltage drop $V_\mathrm{SD}$
across the QPC, plotted with fits to BCS theory (black lines).} 
    \label{fig:mar}
\end{figure}

Having identified superconducting transport signatures in both JJs, we next characterize the induced superconducting gap, which sets the expected critical current of the SQPC. The SQPC is tuned to the onset of conductance, where the channel transmission is approaching zero. The resulting finite bias spectroscopy (\figreflet{fig:mar}{a}) shows a suppression of the conductance around $V_{\mathrm{SD}} =0$. In this regime, the superconducting gap is extracted from the position of the conductance peaks, yielding $\Delta^* \approx 55\ueV$ (\figreflet{fig:mar}{b}). As the gate voltage $V_\mathrm{G}$ is reduced, the junction enters a higher-carrier-density, higher-transmission regime, where the induced superconducting gap is extracted from the resistance peaks associated with multiple Andreev reflections (MAR). The differential resistance as a function of the DC voltage bias is measured for different mixing chamber temperatures at $V_\mathrm{G}=-0.6\V$~(\figreflet{fig:mar}{c}). The resistance peaks corresponding to the first three orders of MAR processes are highlighted by arrows. Their temperature evolution~(\figreflet{fig:mar}{d}) is used to extract both the induced superconducting gap $\Delta_0^*=76\ueV$ and the superconducting critical temperature $T_\mathrm{c}=550\mK$ by fitting the BCS relation \cite{waldram_superconductivity_2017} $\Delta^*(T)=\Delta^*_0\tanh\left(1.74\sqrt{T_\mathrm{c}/T-1}\right)$. The extracted parameters are similar in two different devices (cf. \SupSecMar{}) and in agreement with reported values for PtSiGe JJs \cite{hinderling_direct_2024,tosato_hard_2023}. The varying gap size is also consistent with previous studies reporting an increase in the induced gap with increasing channel transmission and carrier density~\cite{van_loo_electrostatic_2023,kjaergaard_quantized_2016}, highlighting the dependence of the induced superconducting properties on the electrostatic tuning of the semiconductor channel.

\begin{figure*}
    \centering
    \includegraphics[width=\textwidth]{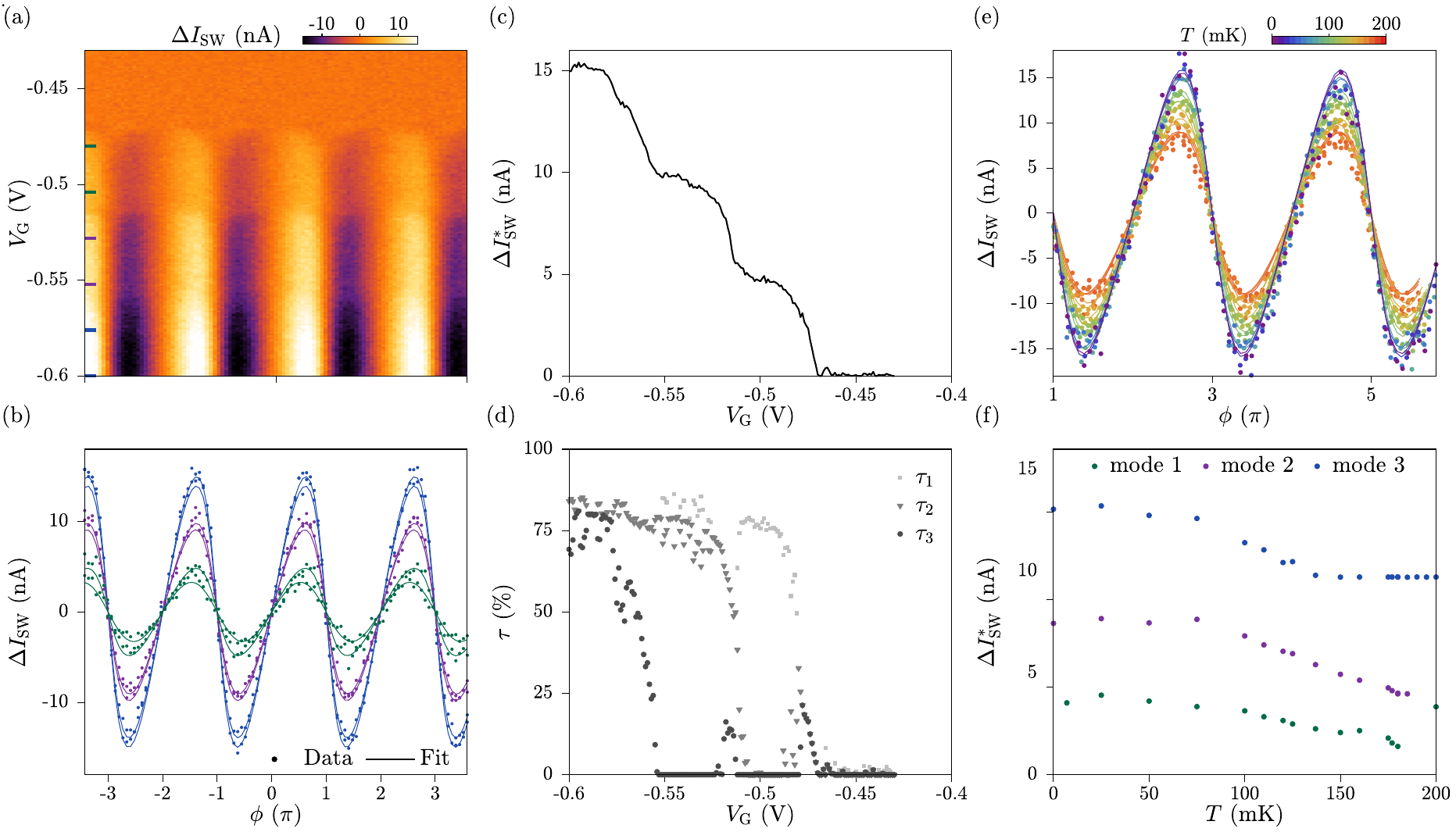}
    \caption{Current-phase relation of device A.  
    \figletter{a}~Average switching current of the device in the SQUID configuration with the switching current of the reference junction subtracted. Markers on the left edge of the plot indicate the position of the line cuts in panel b. \figletter{b}~Line cuts of panel a for different plunger gate voltages $V_\mathrm{G}$ highlighting the amplitude quantization. The data is fitted assuming three modes with variable transmission. \figletter{c}~Fitted amplitude of the SQUID oscillations as a function of plunger gate voltage $V_\mathrm{G}$. \figletter{d}~Fitted transmission of three individual modes. The assignment of transmission to each mode is determined by the maximum value. 
    \figletter{e}~Temperature dependence of the CPR at $V_\mathrm{G}=-0.65\V$ (corresponding to the third conductance plateau of the QPC).  \figletter{f}~Extracted temperature dependence of the SQUID oscillation amplitude on each plateau.}
    \label{fig:CPR}
\end{figure*}

\subsection{Gate-Tunable Switching Current and Current-Phase Relation}
To probe the current–phase relation (CPR) of the SQPC, both JJs are accumulated and the device is operated as a SQUID. The switching current of the reference junction greatly exceeds that of the SQPC, making the SQUID strongly asymmetric and keeping the phase across the reference junction approximately fixed. Instead, the phase across the SQPC varies allowing its CPR to be probed through the modulation of the SQUID switching current. We quantify this modulation by subtracting the switching current of the reference junction, measured with the SQPC depleted at $V_\mathrm{G}=-420\mV$, from the switching current measured at each gate voltage and flux value (\figreflet{fig:CPR}{a}).

The flux-dependent switching-current modulations evolve systematically with $V_\mathrm{G}$ (\figreflet{fig:CPR}{b}). At the most negative gate voltages, the CPR exhibits pronounced forward skewness. As $V_\mathrm{G}$ increases, both the modulation amplitude and the skewness decrease. To quantify the gate dependence of the supercurrent, we extract $\Delta I_\mathrm{SW}^*$ as half the peak-to-dip amplitude of the flux modulation at each gate voltage (\figreflet{fig:CPR}{c}). Three distinct plateaus emerge, closely matching the conductance plateaus measured in the normal state. 

Then, we fit the CPR measurements using the short-junction model.
At zero temperature, the supercurrent carried by a single Andreev bound state (ABS) is given by~\cite{beenakker_josephson_1991}
\begin{equation}
I_{\mathrm{ABS}}(\varphi)=
\frac{e\Delta}{2\hbar}
\frac{\tau\sin\varphi}
{\sqrt{1-\tau\sin^2(\varphi/2)}} ,
\label{Eq:ABS}
\end{equation}
where $\Delta$ is the superconducting gap, $\tau$ is the channel transmission and $\varphi$ is the superconducting phase difference. The total supercurrent is obtained by summing the contributions of the occupied ABSs, each characterized by a transmission parameter $\tau_n$. We consider three modes across the full gate-voltage range and include a gate-independent scaling factor for the oscillation amplitude. The fitted $\tau_n$ values, sorted by magnitude, are shown in \figreflet{fig:CPR}{d}. As the gate voltage is decreased, the effective transmission of the existing modes increases towards a maximum value, while each new switching-current plateau marks the onset of an additional mode.

The temperature dependence of the switching current and CPR are further investigated~(\figreflet{fig:CPR}{e,f}). A progressive reduction of both the switching current and CPR skewness with increasing temperature is observed. The temperature dependence is particularly pronounced in the single-mode regime, where the signal is rapidly suppressed, whereas the three-mode regime appears more robust against thermal fluctuations. Measurements are limited below $200~\mK$, above which our acquisition method no longer allows an unambiguous determination of the switching current.

At base temperature, the observed forward skewness of the CPR is characteristic of highly transmitting channels. The gate-voltage dependence is consistent with the progressive depletion of conducting modes and changes in their effective transmission.
Furthermore, the correlation of $\Delta I_\mathrm{SW}^*$ and the conductance establishes a direct link between the electrostatic occupation of individual transport modes, identified through normal-state conductance quantization, and the superconducting response. Therefore, the stepwise evolution of $\Delta I_\mathrm{SW}^*$ demonstrates that the supercurrent is controlled by opening and closing successive modes in the SQPC.

The measured switching-current steps are approximately $5\nA$, smaller than the ideal short-junction value of $e\Delta/\hbar \approx 18\nA$ expected for a perfectly transmitting mode, using the measured superconducting gap. Similar discrepancies between predicted and measured critical-current steps have been reported in previous studies~\cite{takayanagi_observation_1995, bauch_correlated_2005, abay_quantized_2013, hendrickx_ballistic_2019, mikheev_clean_2023, irie_josephson_2014} and can arise from several mechanisms. Finite channel transmission reduces the supercurrent below the ideal limit, with a nonlinear dependence on $\tau$~\cite{takayanagi_andreev_1995}. In particular, $\tau=0.8$ reduces the expected current step to approximately $55\%$ of its ideal value, bringing the theoretical estimate closer to the experimentally observed magnitude.

Thermal effects can further suppress the supercurrent~\cite{beenakker_superconducting_1992}. At finite temperature, the occupation of excited Andreev bound states (ABSs), whose supercurrent contribution opposes that of the ground state, reduces the net current. Consistently, $\Delta I^*_\mathrm{SW}$ decreases with increasing temperature for all three gate-voltage configurations, as shown in \figreflet{fig:CPR}{f}. This behavior is consistent with thermal occupation of excited ABSs, which also reduces the amplitude and skewness of the current–phase relation (CPR).

\begin{figure*}
    \centering
    \includegraphics[width=\textwidth]{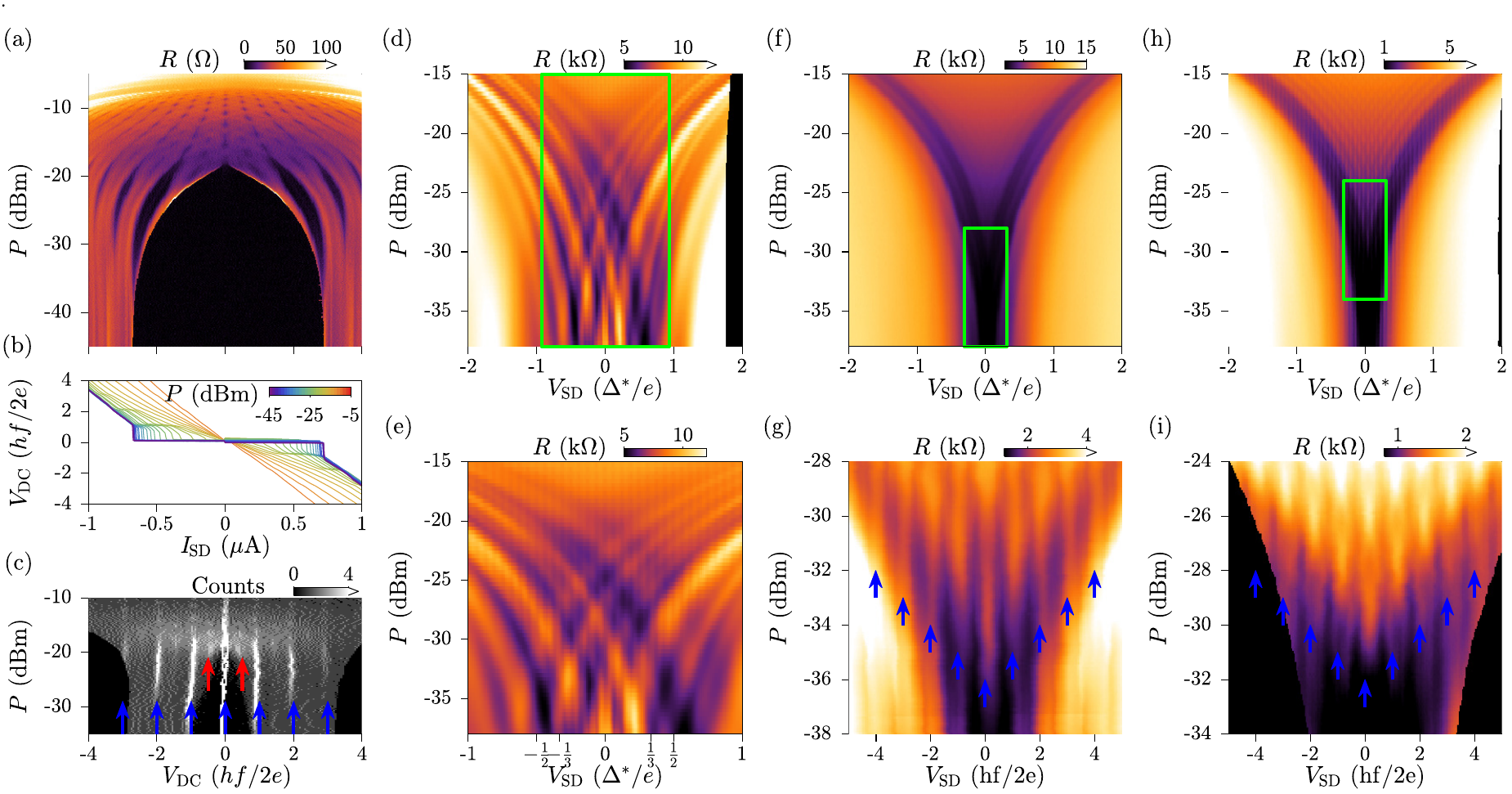}
    \caption{Microwave irradiation of individual JJs of device B. 
    \figletter{a}~Differential resistance of the reference junction in dependence of DC-bias current and applied microwave power at $f=2.25\GHz$. \figletter{b}~DC voltage measured across the reference junction for various microwave powers. \figletter{c}~Histogram of the measured DC voltage across the reference junction for various microwave powers. The integer multiples of $hf/2e$ are highlighted by the blue arrows. The red arrows indicate the half integer steps. \figletter{d, f, h} Differential resistance of the SQPC as a function of DC-bias voltage in units of the superconducting gap and microwave power at fixed $f=2.25\GHz$ for gate voltages $V_\mathrm{G}=-570\mV$ (d),   $V_\mathrm{G}=-580\mV$ (f) and  $V_\mathrm{G}=-650\mV$ (h). The regions depicted in panels e, g, i are marked by the green boxes. \figletter{e, g, i} Zoom of panels d, f, h. The blue arrows indicate the integer multiples of $hf/2e$.}
    \label{fig:RF}
\end{figure*}

Additionally, the measurement of the switching current is an underestimate of the intrinsic critical current because of the junctions' switching dynamics. The switching current of the SQUID is largely determined by the switching dynamics of the reference junction, due to the asymmetry of the SQUID. Although switching-current statistics of the reference junction were acquired, conventional models for underdamped Josephson junctions~\cite{li_quantitative_2002, della_rocca_measurement_2007, Lee_electrically_2011, murphy_universal_2013, haxell_measurements_2023} do not adequately describe the observed behavior, preventing a reliable quantitative correction, as discussed in \supmat{} \SupRefFigStat{}. The measured step heights therefore reflect the combined influence of finite transmission, thermal ABS occupation, and switching dynamics, rather than the intrinsic critical current alone. Despite these limitations, the measured plateaus retain the channel-by-channel evolution expected from conductance quantization, indicating that the switching-current modulation captures the progressive contribution of individual transport modes. 

The unconstrained three-mode fits reproduce the mode occupation inferred independently from normal-state conductance quantization (\figreflet{fig:CPR}{d}). Specifically, one mode contributes significantly on the first conductance plateau, two on the second, and three on the third. Within each plateau, the effective transmission of the corresponding modes increases as the plateau develops. This agreement was obtained without constraining the number, ordering, or gate-voltage dependence of the fitted transmissions. Despite the larger uncertainty associated with nearly closed modes, the fit therefore captures both the sequential opening of transport modes and their continuous transmission tuning by the gate.
This extends previous studies of SNS junctions, in which gate control primarily modulated the number of occupied modes. Within the adopted model, the CPR skewness provides a means of extracting effective channel transmissions. However, since the CPR shape can be affected by mechanisms beyond the intrinsic channel transmission, including thermal occupation of excited ABSs, the extracted values should be regarded as lower bounds on the actual channel transmissions.

\subsection{Microwave-Driven Supercurrent Dynamics}
After discussing the equilibrium phase dependence of the supercurrent, complementary information is obtained from the microwave response of the junctions. We therefore study another device (device B) with a SQPC  length $400\nm$ and gate width $80\nm$. The gate S1 is connected to a cryogenic bias-T, 
which allows to combine DC gate voltage $V_\mathrm{S1}$ and a microwave drive tone. While the RF signal is predominantly affecting the SQPC, it also has a weaker influence on the reference junction. The differential resistance of the reference junction as a function of the bias current and the microwave power for a frequency $f=2.25\GHz$ is plotted in~\figreflet{fig:RF}{a}. As a function of the microwave power, the characteristic fan-like pattern of a driven Josephson junction emerges. The voltage drop across the junction is quantized exhibiting voltage plateaus associated with individual Shapiro steps (\figreflet{fig:RF}{b}). In addition to the integer multiples of $hf/2e$ (blue arrows), we observe half-integer plateaus for powers above $-20\dBm$ (red arrows in~\figreflet{fig:RF}{c}). Previous observations of half-integer Shapiro steps have been attributed to highly transparent, near-ballistic transport channels~\cite{ueda_evidence_2020}. Their observation provides further evidence of the high transparency of the SNS interfaces.
To investigate the microwave response of the SQPC and gain insight into the transport processes governing its nonequilibrium dynamics, we measure its resistance under microwave irradiation as a function of both applied power and gate voltage. For different gate voltages $V_\mathrm{G}$, the power dependence of the resistance varies significantly (\figreflet{fig:RF}{d-i}). At $V_\mathrm{G}=-570\mV$, corresponding to the slope of the first conductance plateau, multiple overlapping V-shaped features emerge (\figreflet{fig:RF}{d}), associated with the power-dependent splitting of  MAR features (\figreflet{fig:RF}{e}). Previous experiments have linked modulations in these MAR features to a modified Tien-Gordon equation~\cite{carrad_photon-assisted_2022}. For gate configurations matching the first and second conductance plateau of the SQPC, these MAR features shift to bias voltages beyond the measured range and a power dependence at zero bias dominates~(\figreflet{fig:RF}{f,g}). For small bias voltages, Shapiro replicas are observed as modulations of the resistance occur as multiples of the applied frequency $hf/2e$ (\figreflet{fig:RF}{h,i}). This gate dependence of the overall response highlights the role of channel occupation and transmission in shaping its nonequilibrium transport. 

\begin{figure}
    \centering
    \includegraphics[width=\columnwidth]{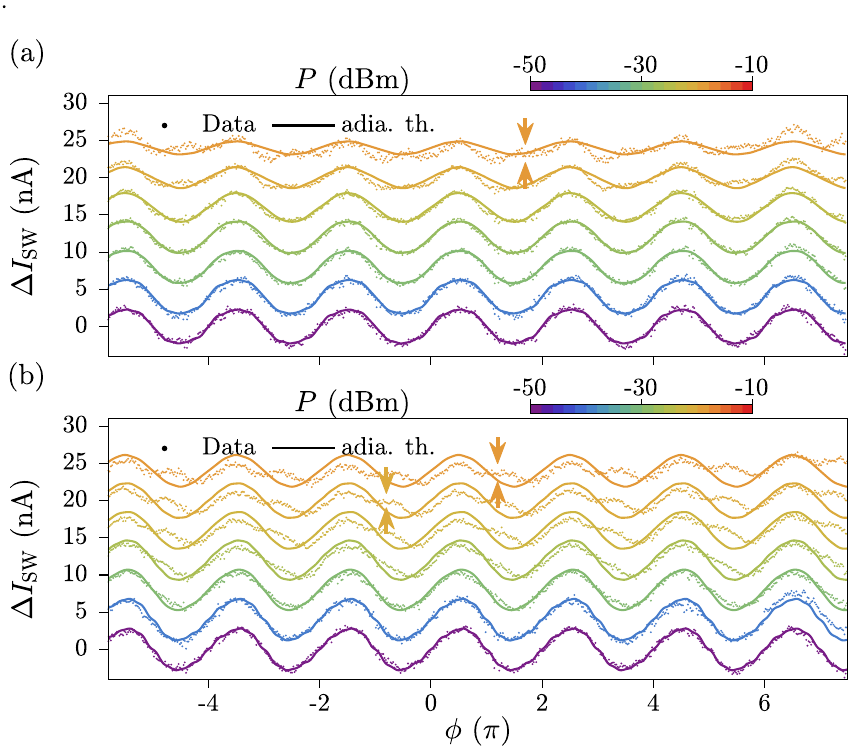}
    \caption{CPR response to microwave irradiation.  
    \figletter{a,b}~Average switching current of the device in the SQUID configuration with the switching current of the reference junction subtracted under microwave radiation for various powers at $f=2\GHz$ (a) and $f=11.5\GHz$ (b). The data is fitted by the adiabatic theory. Deviations are indicated by arrows.}
    \label{fig:RF_CPR}
\end{figure}

The resistance measurements reveal a clear microwave response of the SQPC but do not directly reveal how irradiation affects the ABS occupation and, consequently, the CPR. We therefore investigate the CPR in the power range where the SQPC responds to irradiation while the reference junction maintains a switching current above $0.5\uA$, preserving sufficient SQUID asymmetry for reliable CPR measurements. With increasing microwave power, the CPR amplitude decreases and its skewness evolves, as shown for $f=2~\GHz$ in \figreflet{fig:RF_CPR}{a}. Above $P=-20~\dBm$, the CPR becomes increasingly forward-skewed, while additional features emerge near $\phi=\pi$ at the highest applied power, $P=-18~\dBm$. To investigate the frequency dependence of these effects, we measured the CPR at $f=11.5~\GHz$, adjusting the power range to account for the frequency dependence of microwave coupling and attenuation (\figreflet{fig:RF_CPR}{b}). Compared to the lower-frequency measurements, the CPR exhibits more significant variations from its characteristic forward-skewed shape, some of which are highlighted by arrows. Moreover, comparison of the two highest-power traces reveals opposite trends in the evolution of the CPR skewness at the two excitation frequencies.

The microwave-induced changes can be partly understood within the adiabatic approximation, which predicts a reduction in amplitude and depending on the skewness of the equilibrium measurement an induced change of skewness. Within the adiabatic approximation, the drive is described as a time-dependent modulation of the superconducting phase, $\phi(t)=\phi_0+2\alpha\cos(2\pi ft)$, where $\alpha=eV_\mathrm{MW}/hf$ characterizes the effective microwave coupling~\cite{bergeret_supercurrent_2011,dou_microwave_2021,haxell_microwave-induced_2023}. Expressing the equilibrium CPR in terms of its harmonics, with amplitudes $I_n$, the time-averaged supercurrent is predicted to follow $I_\mathrm{ad}=\sum_n J_0(2n\alpha)I_n\sin(n\phi)$, where $J_0$ is the zeroth-order Bessel function of the first kind. This model reproduces the measured reduction in CPR amplitude but does not account for the observed evolution of the skewness or the emergence of additional features. This suggests that non-equilibrium ABS occupations play an important role~\cite{virtanen_theory_2010,fuechsle_effect_2009}. Microwave absorption or emission can induce resonant transitions when the photon energy $hf$, or a multiple thereof in multiphoton processes, matches the energy separation between ABSs or enables transitions between an ABS and the continuum.

\section{Conclusions}
In conclusion, we investigated a gate-defined superconducting quantum point contact in a \GeSiGe{} heterostructure and demonstrated control of superconducting transport in the few-mode regime. The observation of conductance quantization, quantized switching-current plateaus, and strongly skewed current-phase relations provides direct evidence of superconducting transport mediated by a controlled number of highly transmitting channels. Quantitative analysis of the CPR reveals continuous tuning of individual channel effective transmissions. Under microwave irradiation, distinct transport regimes dominated by photon-assisted Andreev processes and phase-locking effects are determined. The observed microwave-induced modifications of the CPR further highlight the rich non-equilibrium dynamics of Andreev bound states in these devices. Together, our results establish \GeSiGe{} SQPCs as a promising platform for studying mode-resolved superconducting transport and for engineering microwave-driven superconducting-semiconducting quantum circuits.

\begin{acknowledgments}
The authors thank  the Cleanroom Operations Team of the Binnig and Rohrer Nanotechnology Center (BRNC) for their help and support.
\paragraph*{\bf Funding}
This work was supported by the Swiss National Science Foundation (SNSF) under grant no. 200021-227835. A.T.\ acknowledges support from the INSPIRE Potentials SPIN Master’s fellowship. 
\paragraph*{\bf Author contributions}
A.T. and I.S.\ designed the samples, performed the experiments and analysed the data. I.S.\ fabricated the samples.
A.T., I.S.\ and F.N.\ interpreted the data with contributions from I.G.I.\ and T.T.. S.W.B.\ grew the heterostructures. 
F.N.\ conceived the experiment and supervised the project.
A.T.\ and I.S.\ wrote the manuscript with input from F.N. and with contributions from all authors.
\paragraph*{\bf Competing interests}
The authors declare no competing interests.
\end{acknowledgments}

\section*{Materials and Fabrication}
The Ge/SiGe heterostructure hosting the superconducting quantum point contact (SQPC) was grown on a Si substrate by reduced-pressure chemical vapor deposition (RPCVD)~\cite{massai_impact_2024}. The heterostructure consists of a strain-relaxed Ge layer, a graded SiGe buffer, and a compressively strained Ge quantum well embedded between SiGe barriers. The approximately $20\nm$-thick Ge quantum well, located about $48\nm$ below the surface, hosts a two-dimensional hole gas from which the one-dimensional transport channel is electrostatically defined.

The device was fabricated by electron-beam lithography following the procedure described in~\cite{hinderling_direct_2024}. Electrical contacts were defined by Pt evaporation and lift-off. Before metal deposition, the native oxide was removed by briefly dipping the sample in buffered hydrofluoric acid (HF). The deposited Pt was subsequently diffused into the heterostructure by rapid thermal annealing at $350~^\circ\mathrm{C}$, forming platinum germanosilicide (PtSiGe) contacts. A $12\nm$-thick SiO\(_x\) layer was then deposited over the entire sample by atomic layer deposition (ALD), providing electrical insulation between the heterostructure and the gate electrodes. Finally, gate electrodes, consisting of $5\nm$ of Ti and $80\nm$ of Pd, were deposited by evaporation and lift-off.

\section*{Measurement Setup}
Measurements were performed in two dilution refrigerators, each with a base temperature of approximately $10\mK$, with device A measured in one refrigerator and devices B and C in the other. The samples were mounted on QDevil QBoard holders. 
Transport measurements were performed in both voltage- and current-bias configurations. In voltage bias (\figreflet{fig:device}{c,d,e}, \figref{fig:mar}), a DC source–drain voltage from a DC voltage source was combined with a $5\uV$ AC excitation voltage generated by a lock-in amplifier at frequencies below $1~\mathrm{kHz}$. The DC and AC currents were measured using a transimpedance amplifier with a gain of $10^8~\mathrm{V/A}$, while the AC voltage drop was measured using a differential amplifier (gain $1000$). The differential conductance was obtained as $G=I_{\mathrm{AC}}/V_{\mathrm{AC}}$.

For current-biased measurements (\figreflet{fig:device}{f} and \figreflet{fig:RF}{a-c}) a $5\nA$ AC current and DC current were combined, and the resulting AC voltage was measured with a differential  voltage amplifier. For faster determination of the switching current, a ramped current-biased measurement was used. To prevent the electrostatic potential of the device from shifting relative to the gate voltages during current sweeps, the bias was applied using two synchronized sawtooth waveforms of opposite polarity, generated by a waveform generator, on the two ends of the device. The voltage across the device was measured, amplified, and recorded using an oscilloscope. Ramp repetition rates ranged from $133$ to $373~\mathrm{Hz}$, depending on the measurement. Switching-current distributions were acquired by recording the switching event during each individual current ramp and repeating the measurement $10000$ times for each temperature, as shown in \supmat{} \SupRefFigStat{}, whereas averaged traces (16–128 repetitions) were used to extract the CPR measurements in \figref{fig:CPR} and \figref{fig:RF_CPR}.

For microwave experiments, the signal generator was connected to a  cryogenic bias-T via a microwave line with $47~\mathrm{dB}$ of attenuation. Reported microwave powers refer to the signal-generator output.

\bibliographystyle{apsrev4-1-title} 
\bibliography{references.bib}


\makeatletter
\renewcommand\thesection{\mbox{S\arabic{section}}}
\makeatother
\setcounter{section}{0}     

\newcounter{supfigure} \setcounter{supfigure}{0} 
\makeatletter
\renewcommand\thefigure{\mbox{S\arabic{supfigure}}}
\makeatother

\newcounter{suptable} \setcounter{suptable}{0} 
\makeatletter
\renewcommand\thetable{\mbox{S\arabic{suptable}}}
\makeatother

\newenvironment{supfigure}[1][]{\begin{figure}[#1]\addtocounter{supfigure}{1}}{\end{figure}\ignorespacesafterend}
\newenvironment{supfigure*}[1][]{\begin{figure*}[#1]\addtocounter{supfigure}{1}}{\end{figure*}\ignorespacesafterend}
\newenvironment{suptable}[1][]{\begin{table}[#1]\addtocounter{suptable}{1}}{\end{table}\ignorespacesafterend}
\newenvironment{suptable*}[1][]{\begin{table*}[#1]\addtocounter{suptable}{1}}{\end{table*}\ignorespacesafterend}
\newenvironment{suplongtable}[1][]{\addtocounter{suptable}{1}\begin{longtable}[#1]}{\end{longtable}\ignorespacesafterend}

\makeatletter
\renewcommand\theequation{S\arabic{equation}}
\renewcommand\theHequation{S\arabic{equation}} 
\makeatother
\setcounter{equation}{0}

\makeatletter
\newcommand{\faketablecaption}[2]{
\refstepcounter{suptable}
\setlength{\abovecaptionskip}{\smallskipamount}
\setlength{\belowcaptionskip}{\smallskipamount}
\@makecaption{\tablename\ \thetable}{#1}
}
\makeatother

\newcommand{\RefMainFigOne}{Fig.\,\ref{fig:device}} 
\newcommand{\RefMainFigTwo}{Fig.\,\ref{fig:mar}} 
\newcommand{\RefMainFigThree}{Fig.\,\ref{fig:CPR}} 
\newcommand{\RefMainFigFour}{Fig.\,\ref{fig:RF}} 
\newcommand{\RefMainFigFive}{Fig.\,\ref{fig:RF_CPR}} 

\onecolumngrid 
\clearpage
\newpage
{\centering
\large\textbf
{Supplementary information for: \\ A superconducting quantum point contact in Ge} \\ \rule{0pt}{12pt}
}
\twocolumngrid
\section{Additional characterization for the junctions in isolation}
\label{sec:additional_QPCs}
\begin{supfigure}[ht]
    \centering
    \includegraphics[width=8.5cm]{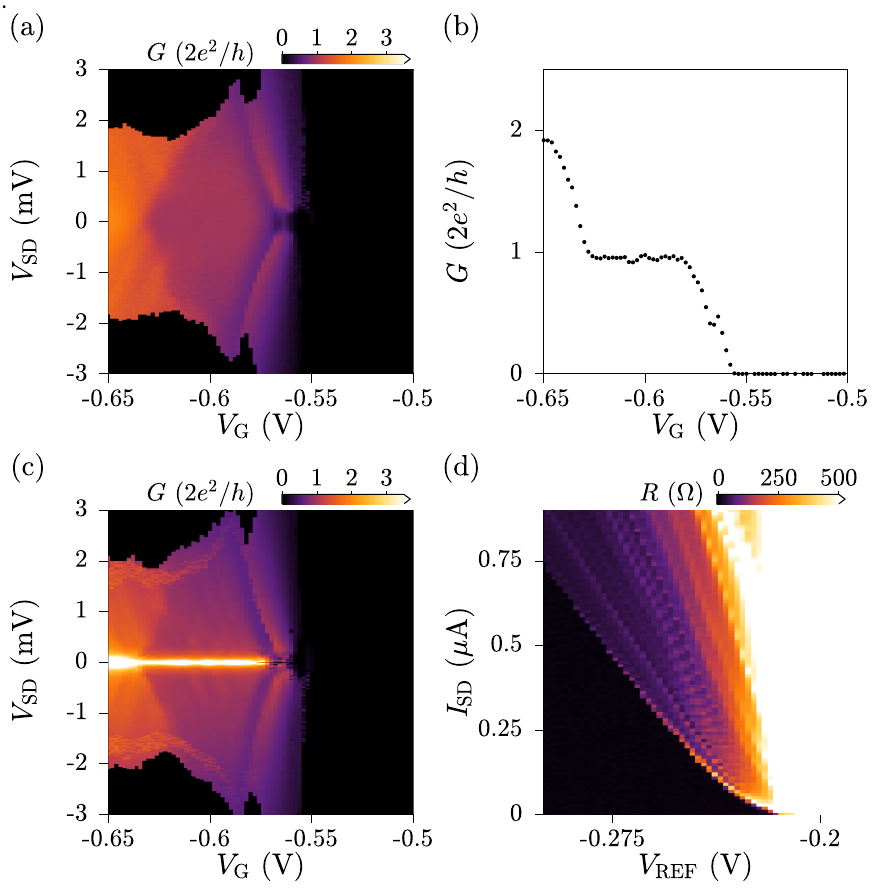}
    \caption{Single JJ characteristics of device B with junction width $80\nm$.
\figletter{a}~Differential conductance as a function of plunger gate voltage $V_\mathrm{G}$ and source-drain bias $V_\mathrm{SD}$ of the SQPC  at $T=700\mK$ and $V_\mathrm{REF}=0\V$. \figletter{b}~Differential conductance of the SQPC  at $T=700\mK$, $V_\mathrm{REF}=0\V$ and $V_\mathrm{SD}=0\V$. \figletter{c}~Differential conductance as a function of plunger gate voltage $V_\mathrm{G}$ and source-drain bias $V_\mathrm{SD}$ of the SQPC  at $T=7\mK$ and $V_\mathrm{REF}=0\V$. \figletter{d}~Differential resistance of the reference junction as a function of bias current $I_\mathrm{SD}$ and gate voltage $V_\mathrm{REF}$ at $V_\mathrm{G}=0\V$.}
    \label{supfig:QPC_devB}
\end{supfigure}
\figref{supfig:QPC_devB} and \figref{supfig:QPC_devC} show measurements equivalent to those presented in \RefMainFigOne (c-f) of the main text for two additional devices with a gate width of $80\nm$. 
All datasets exhibit quantized conductance plateaus in the normal state, a pronounced superconductivity-induced zero-bias conductance enhancement at base temperature, and continuous gate control of the reference junction critical current. For these devices with width $80\nm$, only two conductance plateaus were resolved in the considered voltage regime. The measurements depicted in \figref{supfig:QPC_devB} were obtained for the device used to study the microwave response in \RefMainFigFour{}. For the measurements presented in \RefMainFigFive, device B was cooled down a second time for which the side gates of the QPC were set to $-0.1\V$ to avoid setting the central QPC gate more negative. The corresponding junction characterization is depicted in \figref{supfig:QPC_devB_cooldown2}.

Furthermore, device A used in \RefMainFigOne - \RefMainFigThree{} was cooled down an additional time to measure the temperature dependence of the CPR depicted in \RefMainFigThree (e,f). A small shift in the gate voltage was observed likely due to the hysteretic behavior of the material. Therefore, the voltages chosen to have one, two or three transport modes for the measurements of \RefMainFigThree (f) were adjusted accordingly. 

Overall, the measurements indicate that the transport characteristics of both the superconducting quantum point contact and the reference junction are reproducible across devices and thermal cycles.

\begin{supfigure}[ht]
    \centering
    \includegraphics[width=8.5cm]{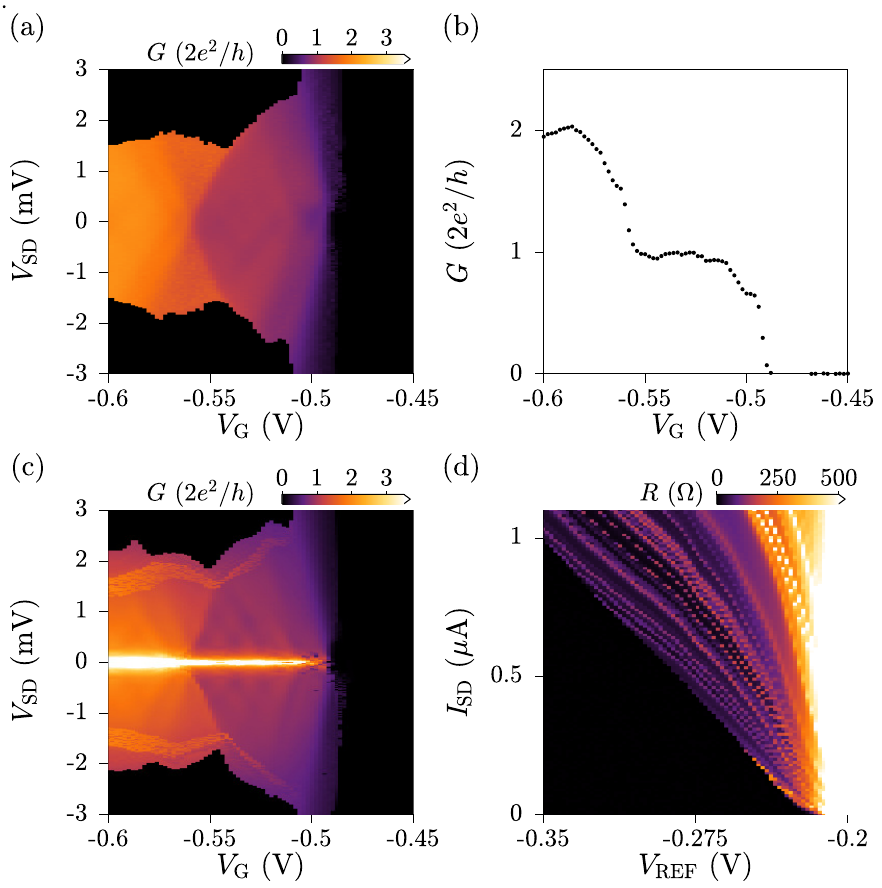}
    \caption{Single JJ characteristics of device C with junction width $80\nm$.  \figletter{a}~Differential conductance as a function of plunger gate voltage $V_\mathrm{G}$ and source-drain bias $V_\mathrm{SD}$ of the SQPC  at $T=700\mK$ and $V_\mathrm{REF}=0\V$. \figletter{b}~Differential conductance of the SQPC  at $T=700\mK$, $V_\mathrm{REF}=0\V$ and $V_\mathrm{SD}=0\V$. \figletter{c}~Differential conductance as a function of plunger gate voltage $V_\mathrm{G}$ and source-drain bias $V_\mathrm{SD}$ of the SQPC  at $T=7\mK$ and $V_\mathrm{REF}=0\V$. \figletter{d}~Differential resistance of the reference junction as a function of bias current $I_\mathrm{SD}$ and gate voltage $V_\mathrm{REF}$ at $V_\mathrm{G}=0\V$.}
    \label{supfig:QPC_devC}
\end{supfigure}

\begin{supfigure}[ht]
    \centering
    \includegraphics[width=8.5cm]{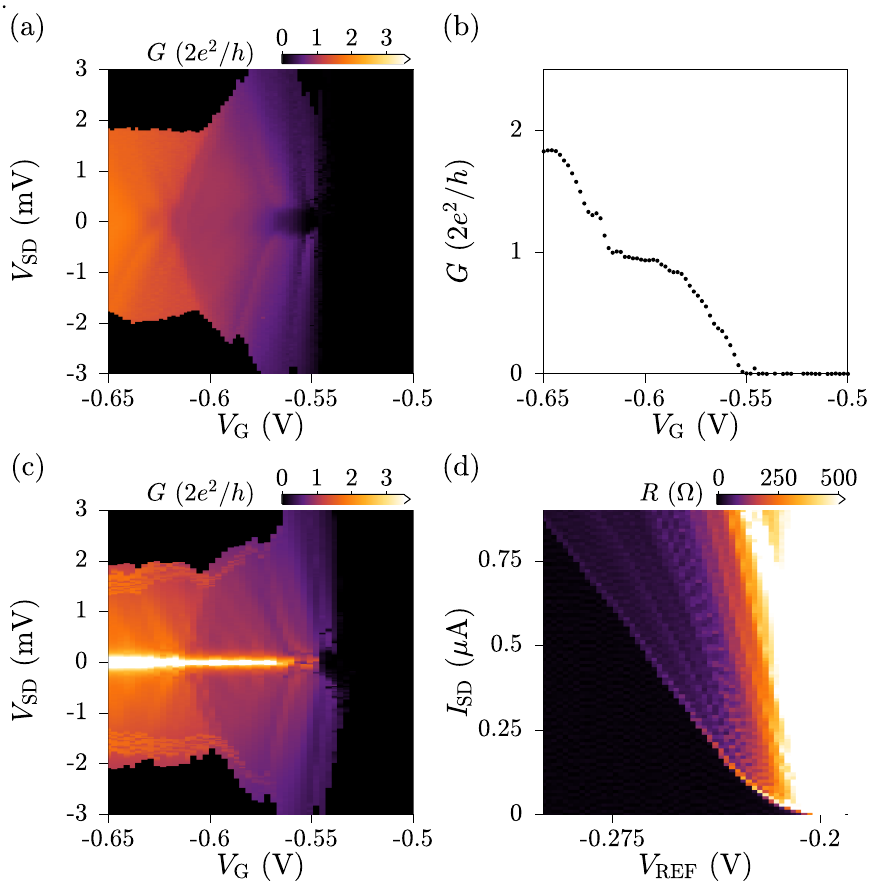}
    \caption{Single JJ characteristics of the second cooldown of device B with junction width $80\nm$. \figletter{a} Differential conductance as a function of plunger gate voltage $V_\mathrm{G}$ and source-drain bias $V_\mathrm{SD}$ of the SQPC  at $T=700\mK$ and $V_\mathrm{REF}=0\V$. \figletter{b}~Differential conductance of the SQPC  at $T=700\mK$, $V_\mathrm{REF}=0\V$ and $V_\mathrm{SD}=0\V$. \figletter{c}~Differential conductance as a function of plunger gate voltage $V_\mathrm{G}$ and source-drain bias $V_\mathrm{SD}$ of the SQPC  at $T=7\mK$ and $V_\mathrm{REF}=0\V$. \figletter{d}~Differential resistance of the reference junction as a function of bias current $I_\mathrm{SD}$ and gate voltage $V_\mathrm{REF}$ at $V_\mathrm{G}=0\V$.}
    \label{supfig:QPC_devB_cooldown2}
\end{supfigure}

\begin{supfigure}[ht]
    \centering
    \includegraphics[width=8.5cm]{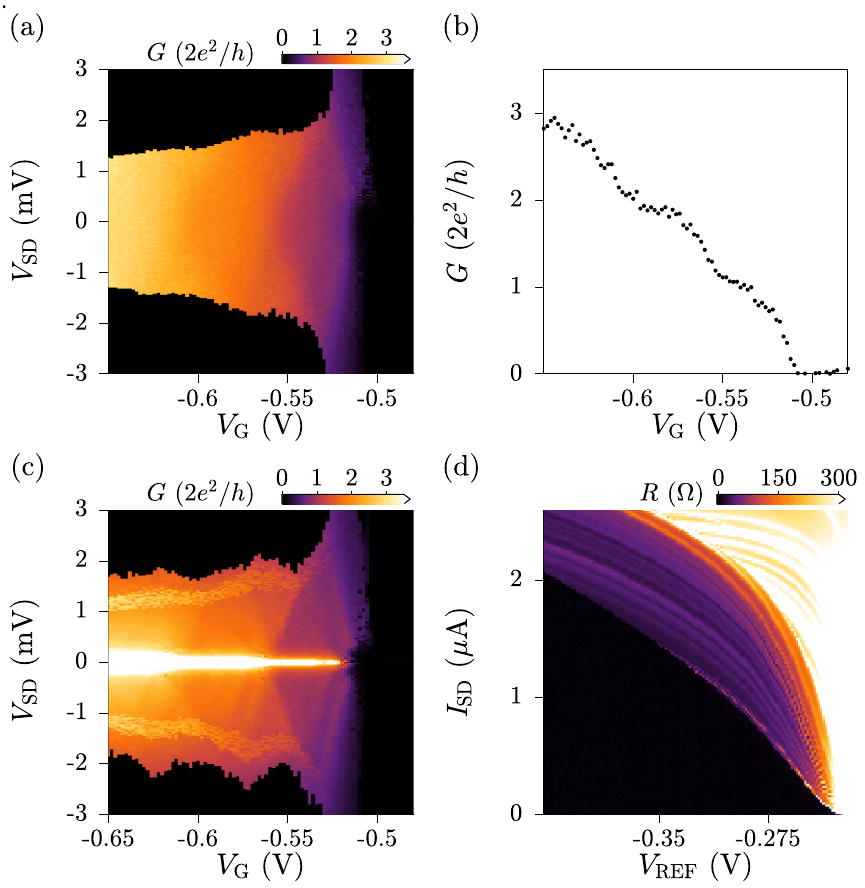}
    \caption{Single JJ characteristics of a different cooldown of device A.
    \figletter{a}~Differential conductance as a function of plunger gate voltage $V_\mathrm{G}$ and source-drain bias $V_\mathrm{SD}$ of the SQPC  at $T=700\mK$ and $V_\mathrm{REF}=0\V$. \figletter{b}~Differential conductance of the SQPC  at $T=700\mK$, $V_\mathrm{REF}=0\V$ and $V_\mathrm{SD}=0\V$. 
     \figletter{c}~Differential conductance as a function of plunger gate voltage $V_\mathrm{G}$ and source-drain bias $V_\mathrm{SD}$ of the SQPC  at $T=7\mK$ and $V_\mathrm{REF}=0\V$. \figletter{d}~Differential resistance of the reference junction as a function of bias current $I_\mathrm{SD}$ and gate voltage $V_\mathrm{REF}$ at $V_\mathrm{G}=0\V$.}
    \label{supfig:QPC_cooldown3}
\end{supfigure}

\FloatBarrier

\section{Multiple Andreev reflections in device C}
\label{sec:sup_mar}
To assess device-to-device variability, we repeat both the tunneling spectroscopy and temperature-dependent MAR measurements on  device C~(\figref{supfig:mar_dev2}). The tunneling spectroscopy reveals a comparable induced superconducting gap of $\Delta ^* \approx 30\ueV$, while the temperature evolution of the MAR features yields values of $\Delta ^* =82\ueV$ and $T=560\mK$ in good agreement with those obtained for device A. The consistency of these measurements across devices confirms the robustness of the induced superconducting properties in our PtSiGe-based junctions.

\begin{supfigure}[h]
    \centering
    \includegraphics[width=8.5cm]{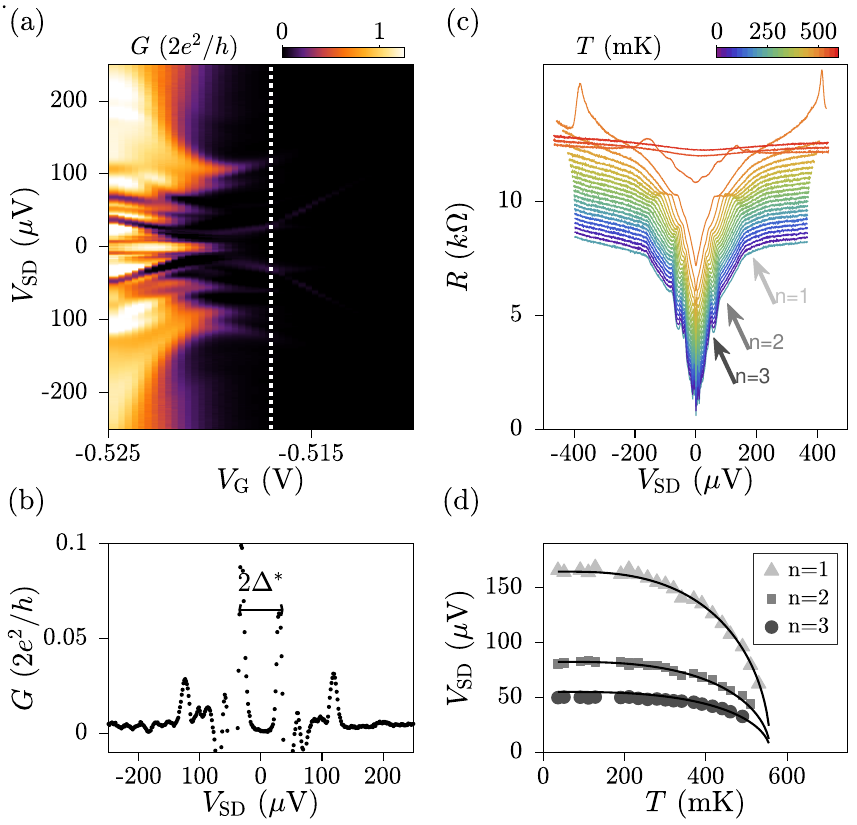}
    \caption{Multiple Andreev reflection of the SQPC of device C.  
    \figletter{a}~Differential conductance of the SQPC at the onset of the first conductance plateau of the SQPC. 
    \figletter{b}~Line cut of panel a at $V_\mathrm{G}=-0.517\V$ (black dots) with marked conductance peaks distance $2\Delta^*$ \figletter{c}~Differential resistance of the SQPC at $V_\mathrm{G}=-0.6\V$ as a function of source-drain voltage $V_\mathrm{SD}$ and temperature $T$. Each temperature curve is offset by $200~\Omega$. The peaks shifting with temperature indicated by arrows correspond to low-order MAR processes. \figletter{d}~Evolution of the three MAR peaks as a function of the temperature and voltage drop $V_\mathrm{SD}$
across the QPC, plotted with fits to BCS theory (black lines). }
    \label{supfig:mar_dev2}
\end{supfigure}

\FloatBarrier
\section{Current-phase relations of additional devices}

\begin{supfigure}
    \centering
    \includegraphics[width=\columnwidth]{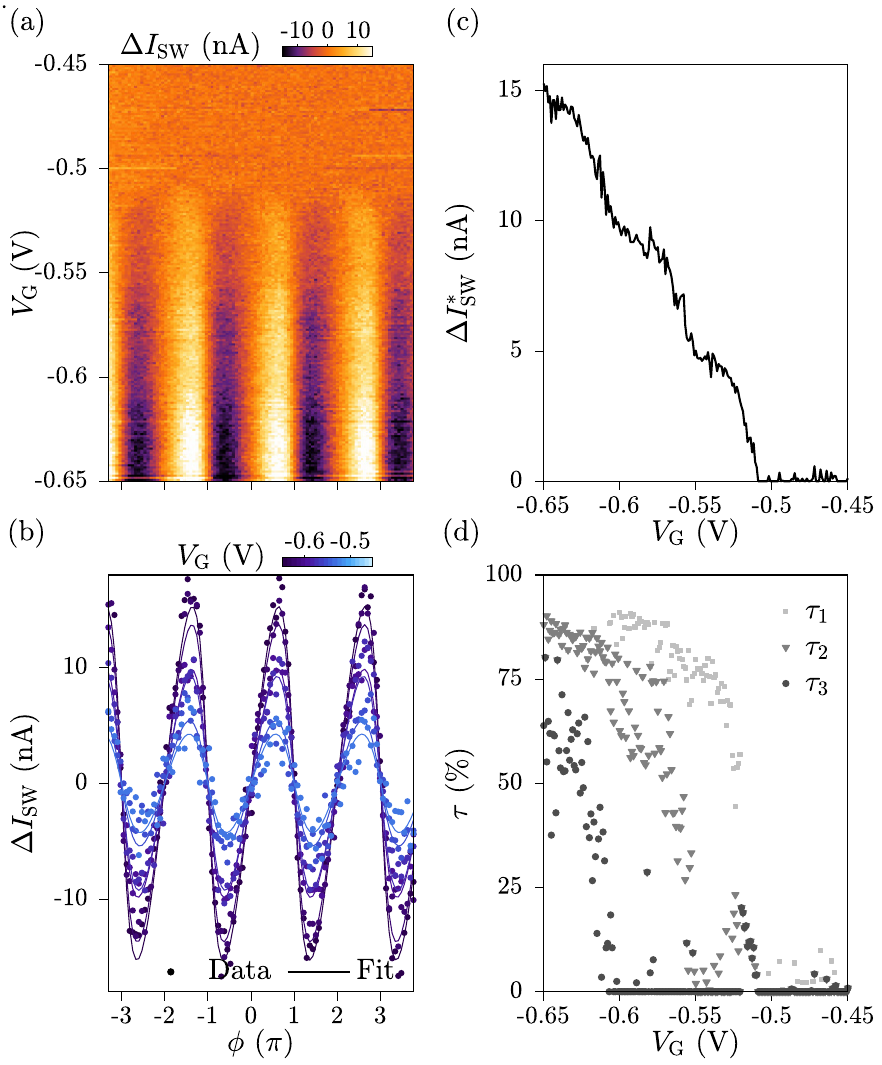}
    \caption{Current-phase relation of device A measured during the second cooldown.  
    \figletter{a}~Average switching current of the device in the SQUID configuration with the switching current of the reference junction subtracted. Markers on the left edge of the plot indicate the position of the line cuts in panel b. \figletter{b}~Line cuts of panel a for different plunger gate voltages $V_\mathrm{G}$ highlighting the amplitude quantization.  \figletter{c}~Fitted amplitude of the SQUID oscillations as a function of plunger gate voltage $V_\mathrm{G}$. \figletter{d}~Fitted transmission of three individual modes. The assignment of transmission to each mode is determined by the maximum value. }
    \label{supfig:CPR_devA_cool2}
\end{supfigure}

\begin{supfigure}
    \centering
\includegraphics[width=\columnwidth]{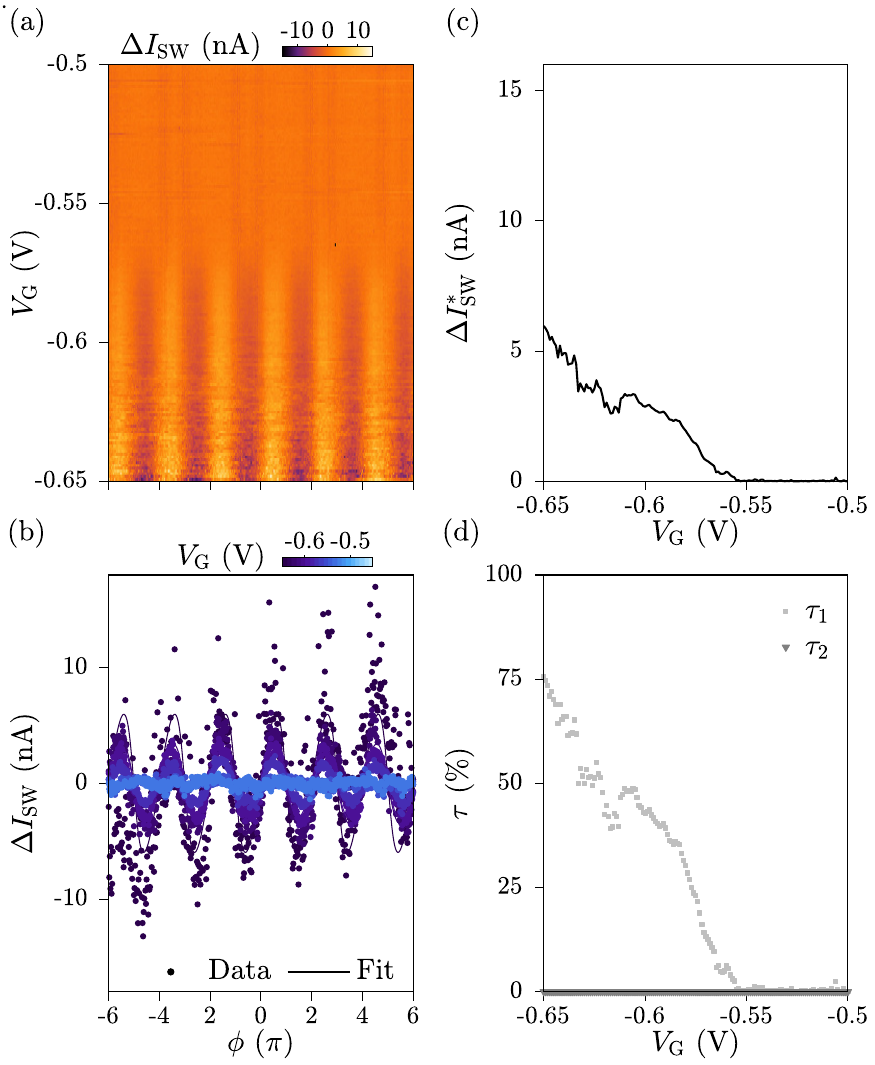}
    \caption{Current-phase relation of device B measured during the second cooldown.  
    \figletter{a}~Average switching current of the device in the SQUID configuration with the switching current of the reference junction subtracted. Markers on the left edge of the plot indicate the position of the line cuts in panel b. \figletter{b}~Line cuts of panel a for different plunger gate voltages $V_\mathrm{G}$ highlighting the amplitude quantization.  \figletter{c}~Fitted amplitude of the SQUID oscillations as a function of plunger gate voltage $V_\mathrm{G}$. \figletter{d}~Fitted transmission of two individual modes. The assignment of transmission to each mode is determined by the maximum value.
    }
    \label{supfig:CPR_devB}
\end{supfigure}

\begin{supfigure}
    \centering
\includegraphics[width=\columnwidth]{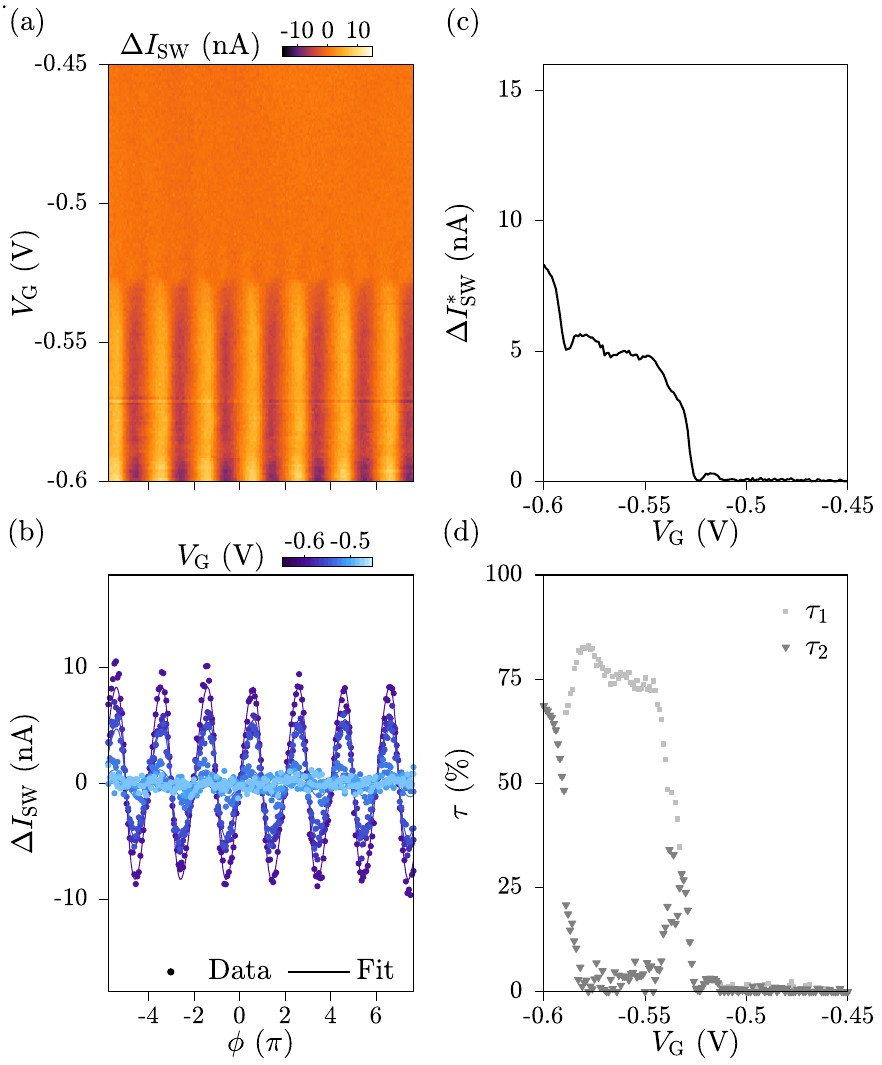}
    \caption{Current-phase relation of device C.  
    \figletter{a}~Average switching current of the device in the SQUID configuration with the switching current of the reference junction subtracted. Markers on the left edge of the plot indicate the position of the line cuts in panel b. \figletter{b}~Line cuts of panel a for different plunger gate voltages $V_\mathrm{G}$ highlighting the amplitude quantization.  \figletter{c}~Fitted amplitude of the SQUID oscillations as a function of plunger gate voltage $V_\mathrm{G}$. \figletter{d}~Fitted transmission of two individual modes. The assignment of transmission to each mode is determined by the maximum value. 
    }
    \label{supfig:CPR_devC}
\end{supfigure}
To assess the reproducibility of the superconducting mode quantization, we performed the same SQUID measurements in the second cooldown of device A and for two additional devices \figrefs{supfig:CPR_devA_cool2}{supfig:CPR_devC}. The extracted SQPC switching current exhibits a clear stepwise evolution with gate voltage, closely mirroring the conductance quantization observed in the normal state. Distinct current plateaus corresponding to the successive occupation of transport modes are resolved. Quantitative analysis of the skewness of the CPR yields transmission probabilities above 75\% for each mode consistent with the behavior observed in \RefMainFigThree{}. For devices B and C, only two separate modes are considered due to the applied gate voltage range. 

\FloatBarrier
\section{Switching statistics}
The switching statistics of the SQUID are dominated by the switching statistics of the reference junction, due to the strong asymmetry of the SQUID. No flux dependence of the SQUID's switching statistics was observed. Therefore, the switching-current distributions (SPDs) of the reference junction and the corresponding escape rates are shown in \figref{supfig:stat}(a,b), respectively. The escape rate $\Gamma(I)$ is calculated as

\begin{equation}
\Gamma(I)= \mathrm{SPD}(I_\mathrm{SW}) \nu \Bigl[ 1 - {\displaystyle \int_0 ^ {I_\mathrm{SW}} \mathrm{SPD}(I)\,dI} \Bigr],
\label{eq from spd}
\end{equation}

where $\nu$ is the current-ramp rate~\cite{brezryadin_stochastic_2012}. The temperature dependence of the mean switching current and the standard deviation of the SPD is summarized in \figref{supfig:stat}(c,d). At the lowest temperatures, the standard deviation is consistent with a crossover regime in which macroscopic quantum tunneling (MQT) and phase diffusion (PD) coexist, with repeated escape and retrapping events preceding the transition to the resistive state. Above approximately $150~\mathrm{mK}$, the standard deviation increases approximately linearly with temperature, indicating a crossover towards thermally activated escape. Monte Carlo simulations based on a model for moderately damped Josephson junctions do not quantitatively reproduce the measured switching distributions~\cite{fenton_monte_2008,longobardi_thermal_2011,haxell_measurements_2023}. This failure suggests that the junction operates in a strongly damped regime, in which repeated escape and retrapping events and phase diffusion become important and cannot be represented adequately by a single effective quality factor. Planar InAs junctions with a normal-state resistance of approximately $150~\Omega$ have previously exhibited $Q\approx7$~\cite{haxell_measurements_2023}. The substantially lower normal-state resistance of our reference junction, approximately $20~\Omega$, together with its small geometric capacitance, is qualitatively consistent with even stronger damping. However, because the relevant high-frequency shunt impedance need not equal the measured normal-state resistance, this comparison does not permit a reliable estimate of Q. Additional contributions from the electromagnetic environment and the coupled dynamics of the asymmetric SQUID may also be relevant. The intrinsic critical current therefore cannot be extracted quantitatively from the switching distributions using the present model.

\begin{widetext}
    \centering
    \includegraphics[width=\textwidth]{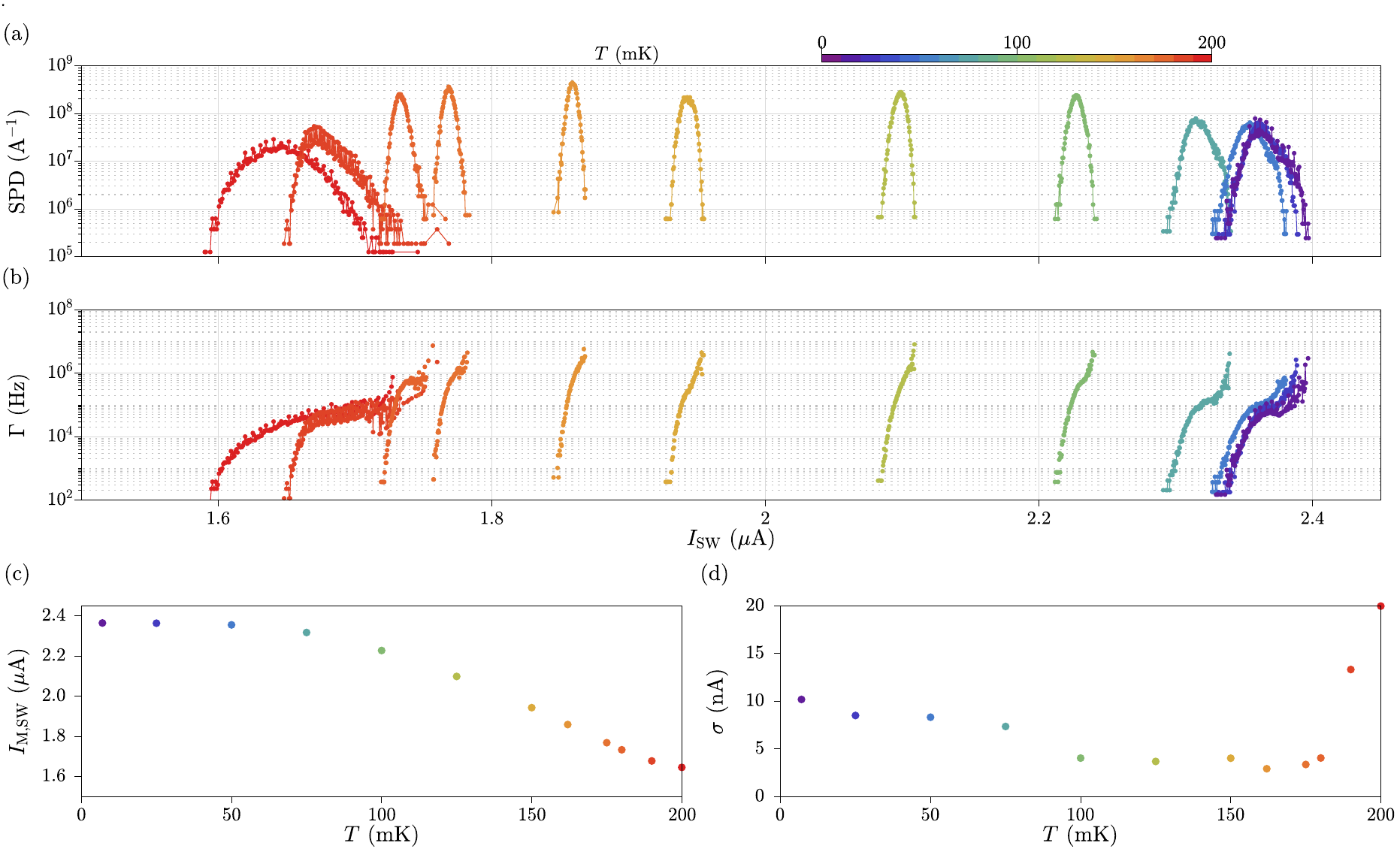}
    \renewcommand{\thesupfigure}{S\arabic{supfigure}}
    \refstepcounter{supfigure}
\makeatletter
\edef\@currentlabel{S\arabic{supfigure}}
\makeatother
\label{supfig:stat}
    \noindent
    \parbox{\textwidth}{Figure \thefigure. Switching statistics of the reference junction of device A.
    \figletter{a} Switching probability distributions (SPDs) measured at different temperatures. 
    \figletter{b} Corresponding escape rates extracted from the SPDs. 
    \figletter{c} Temperature dependence of the mean switching current. 
    \figletter{d} Temperature dependence of the standard deviation of the switching probability distributions.}
    
\end{widetext}

\FloatBarrier
\section{Microwave irradiation at different frequencies}
\parbox{\columnwidth}{%
To further investigate the microwave response of our device, the measurements of \RefMainFigFour{} were repeated with an applied frequency $f=3\GHz$~(\figref{supfig:shapiro}). The reference junction and SQPC exhibit the same qualitative behavior, including integer and half-integer Shapiro steps, MAR-related features, and low-bias Shapiro-step-like resistance modulations. Compared to the data in \RefMainFigFour{}, the features are shifted in microwave power. In addition, a splitting of the shapiro-step-like resistance modulation is observed~(\figreflet{supfig:shapiro}{h,i}).}

\FloatBarrier
\bigskip
\bigskip
\bigskip

\begin{widetext}
    \centering
    \includegraphics[width=\textwidth]{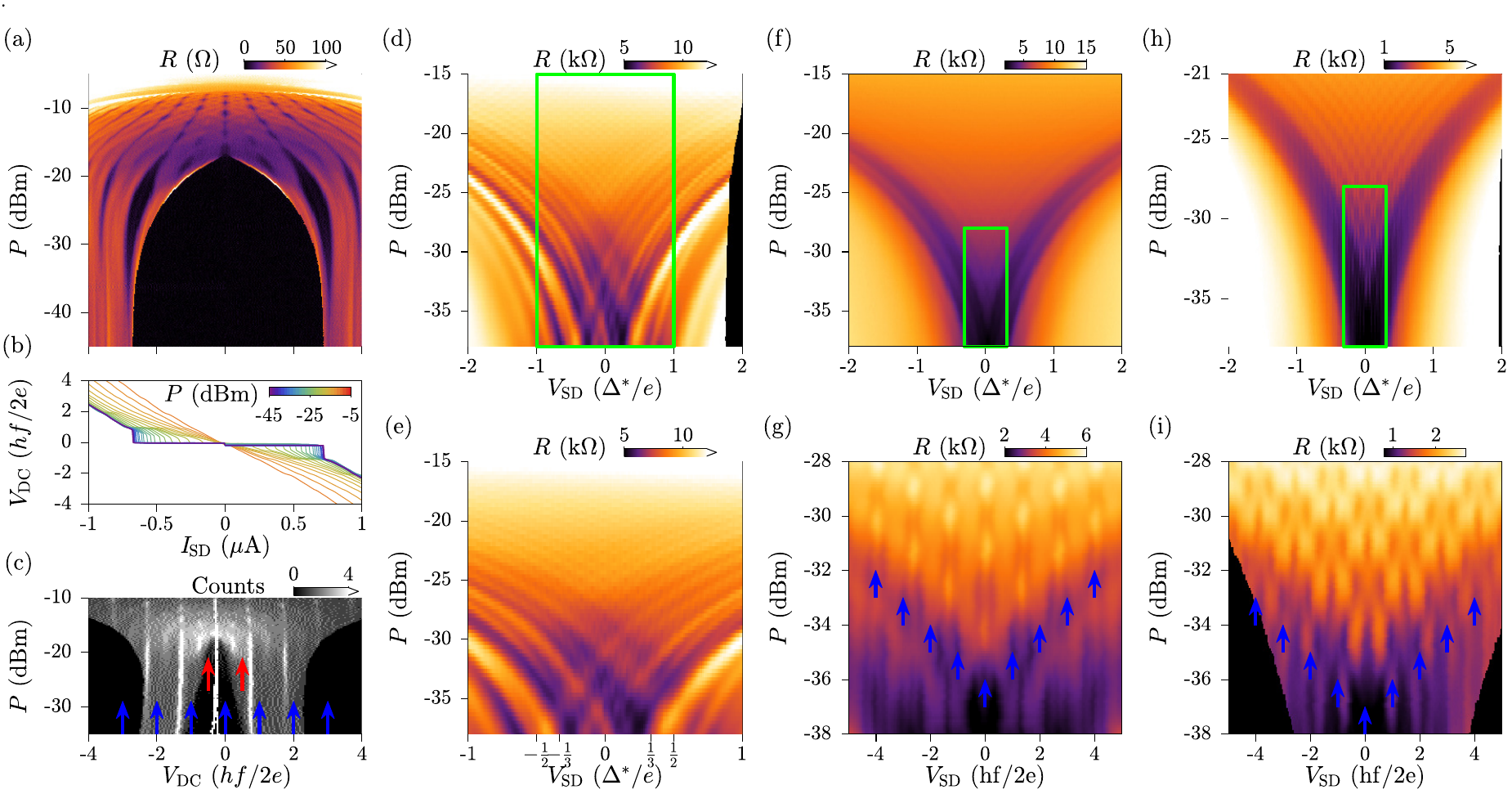}
        \refstepcounter{supfigure}
        \makeatletter
\edef\@currentlabel{S\arabic{supfigure}}
\makeatother
        \noindent
    \parbox{\textwidth}{Figure \thefigure. Microwave irradiation of device B. \figletter{a}~Differential resistance of the reference junction in dependence of DC-bias current and applied microwave power at $f=3\GHz$. 
    \figletter{b}~DC voltage measured across the reference junction for various microwave powers. 
    \figletter{c}~Histogram of the measured DC voltage across the reference junction for various microwave powers. The integer multiples of $hf/2e$ are highlighted by the blue arrows. The red arrows indicate the half integer steps. 
    \figletter{d, f, h} Differential resistance of the SQPC as a function of DC-bias voltage in units of the superconducting gap and microwave power at fixed $f=3\GHz$ for gate voltages $V_\mathrm{G}=-570\mV$ (d),   $V_\mathrm{G}=-580\mV$ (f) and  $V_\mathrm{G}=-650\mV$ (h). The regions depicted in panels e, g, i are marked by the green boxes. \figletter{e, g, i} Zoom of panels d, f, h.}
    
    \label{supfig:shapiro}
\end{widetext}

\end{document}